# THE NEED FOR ADEQUATE SAMPLING IN A WELL-FUNCTIONING MARKET SURVEILLANCE SYSTEM

**Ivan Hendrikx[a] and Nikola Tuneski[b]**

[a]Market surveillance expert, ESTH, Zonhoven, Belgium, hendrikx.ivan@gmail.com

[b]Ss. Cyril and Methodius University, Skopje, R. North Macedonia, nikola.tuneski@mf.edu.mk

**ABSTRACT.**

Adequate sampling is essential for the well-functioning of a market surveillance system. As small as possible statistically significant sample size is the main factor that determines the costs of market surveillance actions. This paper studies various possibilities for calculation of the size of the sample with an emphasis on the method based on the binomial distribution. Examples, comparisons, and conclusions are provided.

1. Introduction

In today's global product regulation regimes, it is of importance that the "putting on the market" of products becomes efficient and easy, this is to say the burden of regulatory compliance before putting the product on the market is lessened, when public authorities are getting less involved. This way of accessing the market by manufacturers or in general "economic operators" must be supported by an efficient and effective market surveillance system. This approach of market access is nowadays promoted by authorities, in particular by Market Surveillance Authorities, but also by interested parties like the economic operators.

From a system level point of view, the main objective of the Market Surveillance Authorities is to make sure that the number of non-conforming products on the market is contained. In the ideal case, for seriously dangerous equipment, should be 0 (zero) %.

There have been created some regulatory documents[1] ([1, 2, 3, 4, 5]) worldwide that put forward such a well-functioning market surveillance system.

This paper discusses "the adequate scale" as is required in a well-functioning MS system as it is said in clause 19 of 765/2008/EC:

> "Market surveillance authorities shall perform appropriate checks on the characteristics of products on an adequate scale, by means of documentary checks and, where appropriate, physical and laboratory checks on the basis of adequate samples."

It may be not clear what „adequate scale" mean s.

---

[1] E.g. Regulation 765/2008/EC [1]; CR12/2012 IOSCO [2]; UNECE WP6 Guide [3]; TCB Post Market Surveillance [5]

It is our understanding that Market Surveillance Authorities in countries are free in defining the contents of these terms.

In that situation, the adequate scale and other detailed market surveillance requirements will be defined by only 1 party: the Market Surveillance Authorities.

In the light of total harmonization of regulations, this may induce problems such as that some member countries may require more serious requirements (e.g. larger sampling schemes, and so on). It may also make the Market Surveillance system less transparent hence not so acceptable by economic operators and finally the public.

## 2. The sampling procedure

A sampling procedure is a set of operational requirements and/or instructions relating to taking and constituting a sample ([6]). In this paper we focus on the different ways of establishing the size of the sample for market surveillance needs. In addition, based on the number of non-conforming items in the sample we will give formulas for calculating the relevant conformity indicators and methodology for delivering a conclusion whether or not the real conformity rate is above some limit value. Actual random sampling is a separate problem outside the scope of this paper.

After carefully studying the possibilities offered within the theory of statistics, we can conclude that there are four different methods that can be used for determining the size of the sample for market surveillance purposes. These methods are:
1. Sample size based on binomial distribution:
2. Sample size based on statistical quality control;
3. Sample size based on ISO 2859-1 ([7]);
4. Sample size based on Bayesian statistics.

Bayesian statistics (method 4) is promising, but still on an academic level, not ready for practical usage.

Both, the statistical quality control (method 2), as well as ISO 2859-1 (method 3), are developed for sampling within production, thus for communication between a producer and a supplier and are essentially accepting procedures, i.e., procedures that end with a decision whether a lot should be accepted or not. This approach is not suitable for sampling for market surveillance needs where we are interested in estimating conformity rate and obtaining a conclusion whether it is above some acceptable value or not.

On the other hand, a binomial experiment is a statistical experiment that has the following properties:
- the experiment consists of *n* repeated trials;
- each trial can result in just two possible outcomes (we call one of these outcomes a success and the other, a failure);
- the probability of success is the same on every trial.

It is obvious that sampling within market surveillance follows completely the above definition.

Therefore, we can conclude that the determination of the sample size by means of the binomial distribution can meet the needs of market surveillance in most efficient way, this means to get meaningful results using minimum sample size.

Further on, the size of the sample using binomial distribution can be determined using the two different approaches:
   a. Sample size based on interval estimate of the population proportion;
   b. Sample size based on the power of the test of hypothesis on a population proportion;

Both variants will be presented in the paper developed using the book Statistical Methods for Rates and Proportions by Fleiss et al. ([8]). Other valuable references on the topic are [9] and [10]. At the end of the paper we illustrate the two approaches by examples, make a comparison between them and draw relevant conclusions.

From chronological point of view it is worth pointing out that initial considerations for application of binomial statistics in market surveillance sampling was done by the authors in [11] and [12].

### 2.a. Sample size based on interval estimate of the population proportion

If in a sample of size $n$ we find $d$ non-conforming, then the best statistical estimate of the conformity rate, i.e., of the fraction of conforming items in the whole population (lot) is

$$f = \frac{n-d}{n} = 1 - \frac{d}{n}.$$

In market surveillance we are further interested in so called one sided interval estimate of the population proportion, i.e., in information that the real conformity rate ($f_r$) is bigger than some value with certain pre-assigned level of confidence (LC). The statistical formula is (see [8], page 28, formula (2.17)):

$$f_r > f_L = \frac{(2nf + z_\alpha^2 - 1) - z_\alpha \cdot \sqrt{z_\alpha^2 - (2 + 1/n) + 4f(n + 1 - nf)}}{2(n + z_\alpha^2)},$$

where $\alpha = 1 - \text{LC}$ is the significance level and $z_\alpha$ is a statistical constant with following values:

Table 1. Values of the level of confidence, significance level and z-value.

| LC | 70% | 75% | 80% | 85% | 90% | 95% | 99% |
|---|---|---|---|---|---|---|---|
| α (significance level) | 0.3 | 0.25 | 0.2 | 0.15 | 0.1 | 0.05 | 0.01 |
| $z_\alpha$ | 0.524 | 0.674 | 0.842 | 1.036 | 1.282 | 1.645 | 2.326 |

For market surveillance purposes the level of confidence can be taken $\text{LC} = 80\%$ or less.

If we fix in advance the maximal difference between the real conformity rate ($f_r$) and $f_L$ to be $w$, i.e.,

$$w = \max(f_r - f_L),$$

then the minimal sample size that guarantees a certain statement, with level of confidence LC is (see

[8], page 35, formula (2.29)):

$$n \geq \frac{k \cdot z_\alpha^2}{w^2} + \frac{2}{w} - 2z_\alpha^2 + \frac{z_\alpha + 2}{k}$$

where for the coefficient $k$ and preliminary conformity rate $f_P$ (known from past inspections, experts opinion or obtained by preliminary sampling), assuming $w$ is no greater than 0.6 we have

Table 2. Values the coefficient $k$ for different preliminary conformity rates.

| If preliminary conformity rate ($f_P$) satisfies | Then use |
|---|---|
| $0 \leq f_P < w/2$ | $k = 4w(1-w)$ |
| $w/2 \leq f_P < 0.3$ | $k = 4(f_P + w/2)(1 - f_P - w/2)$ |
| $0.3 \leq f_P \leq 0.7$ | $k = 1$ |
| $0.7 < f_P \leq 1 - w/2$ | $k = 4(f_P - w/2)(1 - f_P + w/2)$ |
| $1 - w/2 < f_P \leq 1$ | $k = 4w(1-w)$ |

If no knowledge of $f_P$ is available, then $k = 1$.

### 2.b. Sample size based on the power of the test of hypothesis on a population proportion

Again, as in the beginning of the previous section, let us assume that in a sample of size $n$ there are $d$ non-conforming items and then the best estimate of the conformity rate is

$$f = \frac{n-d}{n} = 1 - \frac{d}{n}.$$

Depending on the risk level of the inspected product, the corresponding acceptable conformity rate (ACR) is usually set as in the next table.

Table 3. Values of the acceptable conformity rate for different product risks.

| Product risk | Low | Medium | High | Serious |
|---|---|---|---|---|
| Acceptable Conforming Rate (ACR) | 80% | 85% | 95% | 99% |

If the estimated conformity rate is smaller than the acceptable one, i.e., if

$$f < \text{ACR}$$

it is legitimate to ask how far below ACR, $f$ should be in order to declare that the inspected product has conformity rate smaller than the acceptable one. In the theory of statistics this question is referred as to so called test of hypothesis and the answer is that if (see [8], page 28, formula (2.15)):

$$f \leq \text{ACR} - z_\alpha \cdot \sqrt{\frac{\text{ACR}(1 - \text{ACR})}{n}} - \frac{1}{2n} \qquad (1)$$

then the real conformity rate is smaller than the acceptable conformity rate (ACR) with level of confidence LC. Here, $n$ is the sample size and $z_\alpha$ is defined in the previous section. The quantity $1/(2n)$ should be omitted if it is numerically comparable (close to) the number $|f - \text{ACR}|$.

Otherwise, if $f < \text{ACR}$ but (1) is not satisfied, or if $f \geq \text{ACR}$, then we cannot say that the real conformity rate is smaller than the acceptable conformity rate, i.e., the inspected product is conforming.

The value $\alpha = 1 - \text{LC}$ is the significance level introduced in the previous section. In theory of statistics it is also called type I error, or in market surveillance and quality control better known as the producer's risk.

Important additional information is the so called power of the test which is the probability to detect a "bad" product (see [8], page 31, formula (2.21)):

$$\text{power (for } f) \approx P\left[Z \leq \frac{n(\text{ACR} - f) - z_\alpha \cdot \sqrt{n \cdot \text{ACR}(1 - \text{ACR})}}{\sqrt{n \cdot f(1 - f)}}\right],$$

where Z denotes a standard normal random variable.

The complement of the power is denoted by $\beta$, i.e., $\beta = 1 - power$, and it is the so called type II error or in market surveillance and quality control better known as the consumers' risk ("bad" product passes the inspection).

If we fix in advance the producer's risk ($\alpha$) and the consumers' risk ($\beta$), then the sample size which ensures that values of $\alpha$ and $\beta$, is (see [8], page 32, formula (2.22)):

$$n \geq \left[\frac{z_\alpha \cdot \sqrt{f_P(1 - f_P)} + z_\beta \cdot \sqrt{\text{ACR}(1 - \text{ACR})}}{\text{ACR} - f_P}\right]^2,$$

where $f_P$ is the preliminary conformity rate (known from past inspections, experts opinion or obtained by preliminary sampling).

### 3. Example

We are inspecting a product category of medium risk, i.e., the acceptable conformity rate is 85% (ACR=0.85) with level of confidence 80% (LC = 80% and producer's risk $\alpha = 1 - 0.8 = 0.2$). For the interval estimate of the conformity rate we are willing to accept interval of width 0.1 ($w = 0.1$) and for the test of hypothesis we want power of 90% ($\beta = 1 - 0.9 = 0.1$ is the consumers' risk).

Using the formulas from section 2.a and 2.b we receive the following chart for the sample size based on interval estimate of the population proportion and based on the power of the test of hypothesis on a population proportion.

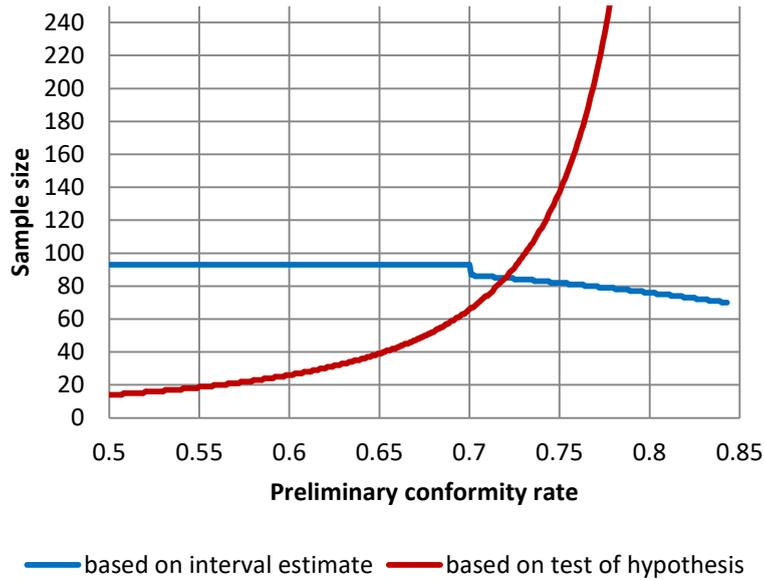

Fig 1. Sample size based on the interval estimate and on test of hypothesis

We can realize that when the preliminary conformity rate is close to the acceptable conformity rate, then the sample size based on the test of hypothesis raises dramatically and in that case sample size based on the interval estimate should be used. Otherwise, if the preliminary conformity rate is far from its acceptable value, then sample size based on the test of hypothesis should be applied. Some numerical values are given in the table below.

Table 4. Sample sizes for different preliminary conformity rates.

| Preliminary conformity rate | 0.5 | 0.6 | 0.65 | 0.7 | 0.75 | 0.8 |
|---|---|---|---|---|---|---|
| Sample size (test of hypothesis) | 14 | 26 | 39 | 66 | 137 | 498 |
| Sample size (interval estimate) | 93 | 93 | 93 | 93 | 82 | 76 |

So, if the preliminary conformity rate is known to be 0.7 ($f_P = 0.7$), it is far from the ACR= 0.85, and so the approach based on the test of hypothesis should be applied. Next figure shows the changes of the power of the test when the sample size varies.

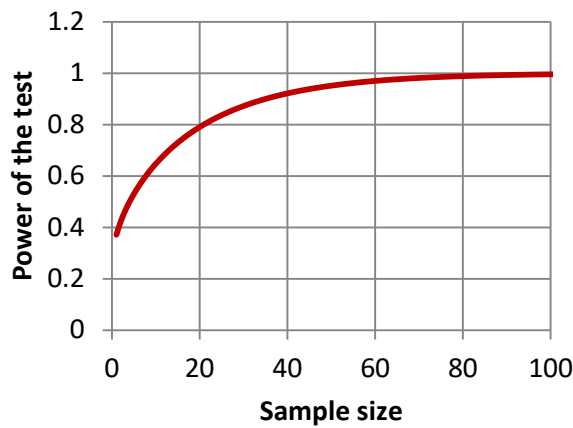

Fig. 2. Power of the test vs. sample size

Notably, the power rises as the sample size rises and a compromise with the costs of inspection is to be made. Some numerical values are given in the table below.

Table 5. Sample sizes for different power of the test.

| Sample size | 13  | 17   | 21  | 27   | 36  | 50   |
|-------------|-----|------|-----|------|-----|------|
| Power       | 0.7 | 0.75 | 0.8 | 0.85 | 0.9 | 0.95 |

In the case when the preliminary conformity rate is known to be 0.8 ($f_P = 0.8$), i.e., it is close to the ACR= 0.85, the approach based on the test of hypothesis should be applied. Next figure shows the changes of the sample size over different widths of the interval estimate. Smaller sample size (meaning lower costs) corresponds to larger width (lower precision of the estimate), so again a compromise should be made between those two.

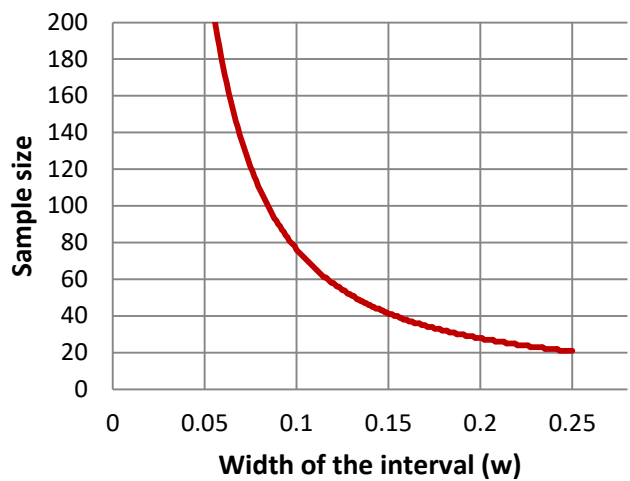

Fig. 3. Sample size vs. width of the interval

Some numerical values are given in the table below.

Table 6. Sample sizes for different widths of the interval estimate.

| Width of the interval estimate | 0.1 | 0.15 | 0.2 |
|--------------------------------|-----|------|-----|
| Sample size                    | 76  | 41   | 28  |

## 4. Conclusions

According to the 'life span' (see the graphic in figure 4) of market surveillance actions, it is important to know when a marker surveillance authority needs to organize a market surveillance action, to bring the CR of the selected products within the acceptable range (e.g. between 75 and 85 %). It has been shown in this paper that with appropriate choice of statistical parameters reasonable low sample sizes can be reached so the MSA can perform what is called preliminary sampling to get an idea of the actual CR rate. In this way, the MSA has been given a method based on statistical evidence when to start the MS action.

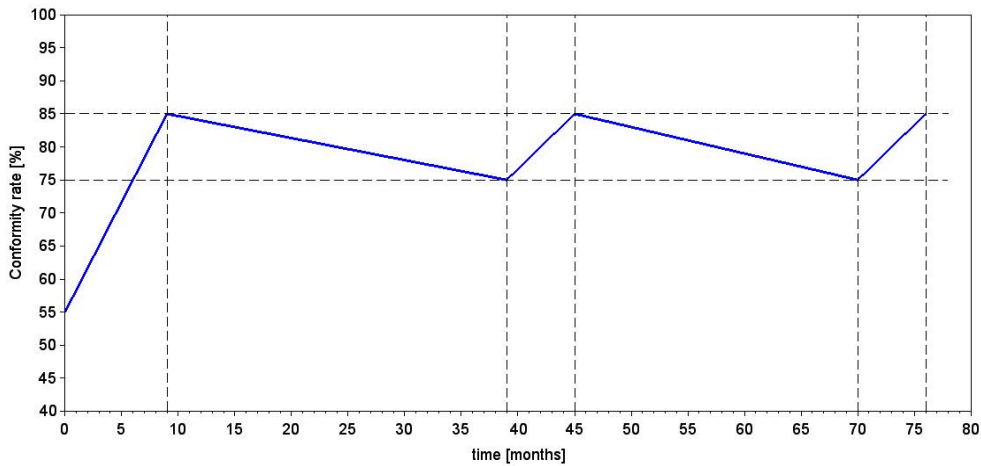

Fig. 4. 'life span' of market surveillance actions

Dynamic simulation of market surveillance actions consisting of creation of a model, its validation and exploration was done by Hendrikx et al. in [13].

REFERENCES

1. "Regulation (EC) No 765/2008 of the European Parliament and of the Council". *Official Journal of the European Union (OJEU)*, 2008.

2. International Organization of Securities Commissions. (2012) *Technological Challenges to Effective Market Surveillance Issues and Regulatory Tools*. [Online]. Available: https://www.iosco.org/library/pubdocs/pdf/IOSCOPD389.pdf

3. Working Party on Regulatory Cooperation and Standardization Policies. (2009). Draft guide to the use of the general market surveillance procedure, [Online]. Available: https://www.unece.org/fileadmin/DAM/trade/wp6/documents/2009/wp6_09_GMS_012E.pdf

4. UNECE Working Party on Technical Harmonization and Standardization Policies. (2001). Recommendations on regulatory cooperation and standardization policies. International Model for Technical Harmonization Based on Good Regulatory Practice for the Preparation, Adoption and Application of Technical Regulations via the Use of International Standards. [Online] Available:
http://www.unece.org/fileadmin/DAM/trade/wp6/Recommendations/Rec_L.pdf.

5. 610077 D01 TCB Post Market Surveillance v06r01. [Online] Available: https://apps.fcc.gov/kdb/GetAttachment.html?id=AWhYotapku%2FSkvVA1wkMAw%3D%3D&desc=610077%20D01%20TCB%20Post%20Market%20Surveillance%20v06r01&tracking_number=20540

6. ISO 3534-2:2006, Statistics - Vocabulary and symbols - Part 2: Applied statistics.

7. ISO 2859-1:1999, Sampling procedures for inspection by attributes - Part 1: Sampling schemes indexed by acceptance quality limit (AQL) for lot-by-lot inspection.

8. Joseph L. Fleiss, Bruce Levin, Myunghee Cho Paik, Statistical Methods for Rates and Proportions, Third Edition, John Wiley & Sons, 2003.

9. Walpole R.E., Myers R.H., Myers S.L., Ye K., Probability & Statistics for Engineering & Scientists, Prentice Hall, 2007.

10. William G. Cochran, Sampling Techniques, 3rd Edition, John Wiley & Sons, 1977.
11. Hendrikx I., Tuneski N., Emerging Issues in Regulations and Standards - Sampling Considerations Within Market Surveillance Actions, *Conformity* Vol. 22 (2009), 23-27.
12. Hendrikx I., Tuneski N., Sampling considerations within Market Surveillance actions, *Proceedings of IEEE Symposium on Product Compliance Engineering*, 26-28 October, 2009, Toronto, Canada, 1-4. DOI 10.1109/PSES.2009.5356011
13. Hendrikx I., Jovanoski B.D., Tuneski N., Dynamic simulations of market surveillance actions, *2016 IEEE Symposium on Product Compliance Engineering* (ISPCE), 16-18 May 2016, Anaheim, CA, USA. DOI: 10.1109/ISPCE.2016.7492846

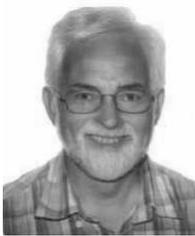

**Mr. Ivan Hendrikx**, can look back at 40 years of experience in **market surveillance, conformity assessment, testing, accreditation and standardization**, demonstrating extensive practice in **policy development, planning and monitoring** of market surveillance activities as well as **coordinating market surveillance institutions**. This includes transposition and implementation of the EU New Legislative Framework (764/2008, 765/2008, 768/2008) as well as New Approach directives, European market surveillance policy formulation related to planning of annual sampling, testing and monitoring of surveillance actions. He can be reached at hendrikx.ivan@gmail.com.

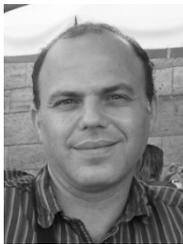

**Mr. Nikola Tuneski**, with his B.Sc. in Mechanical Engineering, M.Sc. in mathematics (probability and statistics) and Ph.D. in pure mathematics (complex analysis), has a rare combination of knowledge and skills in engineering sciences, as well as in applied and pure mathematics. This, together with his proficiency in MATLAB programming led him to a respectful list of 4 published books, 64 original research papers published in well-known journals (23 of them in journals with ISI Thomson Reuters impact factor), participation with a presentation in 54 international conferences and a position of Full Professor at the Ss. Cyril and Methodius University in Skopje. Within the two decades of experience in statistics and sampling, he has been engaged as an expert in a series of projects related to market surveillance, financed among others by EU, Saudi Arabia and PTB (Germany). He can be reached at nikola.tuneski@mf.edu.mk.